\documentclass[prb,preprint]{revtex4-1}
\usepackage{amsmath}
\usepackage{bm}
\usepackage{amsfonts}

\begin{document}

\title{Twelve years before the quantum no-cloning theorem}

\author{Juan Ortigoso}
\email{j.ortigoso@csic.es}
\affiliation{Instituto de Estructura de la Materia, CSIC, Serrano 121, 28006 Madrid, Spain}

\date{January 22, 2018}

\begin{abstract}
The celebrated quantum no-cloning theorem establishes the impossibility of making a perfect copy of an unknown quantum state. The discovery of this important theorem for the field of quantum information is currently dated 1982.
I show here that an article published in 1970 [J. L. Park, Foundations of Physics, 1, 23-33 (1970)] contained an explicit mathematical proof of the impossibility of cloning quantum states. I analyze Park's demonstration in the light of published explanations concerning the genesis of the better-known papers on no-cloning.
\end{abstract}

\maketitle

\section{Introduction}
The no-cloning theorem of quantum mechanics establishes that an arbitrary unknown quantum state cannot be copied. \cite{ike} A modern proof, \cite{merminbook} based on the linearity of quantum mechanics, takes two lines. Suppose that a device can implement a transformation $T$ for copying two orthogonal states $|\psi\rangle$ and 
$|\phi\rangle$ of a qubit:
$T|\psi\rangle |0\rangle=|\psi\rangle |\psi\rangle$  and $T|\phi\rangle |0\rangle=|\phi\rangle |\phi\rangle$, where $|0\rangle$ is the ready state of the target system. It follows, from linearity, that

\begin{equation}
T(a|\psi\rangle + b |\phi\rangle)|0\rangle=a T|\psi\rangle |0\rangle+bT|\phi\rangle |0\rangle=a|\psi\rangle |\psi\rangle+b|\phi\rangle|\phi\rangle \;. 
\label{eqn11}
\end{equation}

\noindent
But if the transformation $T$ can clone arbitrary states, it should give, for any $a$, $b$ values

\begin{equation}
T(a|\psi\rangle + b |\phi\rangle)|0\rangle=(a|\psi\rangle+b|\phi\rangle)(a|\psi\rangle+b|\phi\rangle)=a^{2}|\psi\rangle|\psi\rangle+b^{2}|\phi\rangle |\phi\rangle+ab|\psi\rangle |\phi\rangle+ab |\phi\rangle |\psi\rangle \;,
\label{eqn12}
\end{equation}

\noindent
which is different from Eq. (\ref{eqn11}), unless $a$ or $b$ is zero. 

On the other hand,  the state of a classical system can always be read, in principle, by making appropriate measurements. Thus, classical information like strings of bits, or letters of an alphabet, can be copied, and as a consequence no classical equivalent to the quantum no-cloning theorem exists. This looks paradoxical at first sight since even classical systems must obey quantum mechanics. Nielsen and Chuang, in page 530 of their monograph on quantum computation, \cite{ike} give a simple explanation to the conundrum: ``... the no-cloning theorem does not prevent {\it all} quantum states from being copied, it simply says that nonorthogonal sates cannot be copied... This observation resolves the apparent contradiction between the no-cloning theorem and the ability to copy classical information, for the different states of classical information can be thought of merely as orthogonal quantum states."

It is widely believed that the first versions of the theorem were published in 1982, in two simultaneous and independent articles written by Wootters and Zurek, \cite{WZpaper} and Dieks. \cite{Dpaper} A paper by Milonni and Hardies, \cite{milonni} published in the same issue of Physics Letters as Dieks' paper, also argues that cloning unknown quantum states is impossible: ``Perfect and certain replication of any single photon is impossible," but no explicit mathematical proof was given. 

Asher Peres, a pioneer in quantum information theory, \cite{wperes} wrote a fascinating essay \cite{Ppaper} discussing the events that led to the discovery of the no-cloning theorem. Peres explains that Wootters and Zurek \cite{WZpaper} and Dieks \cite{Dpaper} wrote their articles as a reaction to a paper by Herbert \cite{Fpaper} (the so-called FLASH paper). Peres, one of the referees of the FLASH paper, decided to recommend it for publication knowing that it had to be incorrect, since Herbert's proposal is a strategy to achieve faster than light communication. Peres rightly thought that publication of the erroneous paper would prompt broad interest in the community to find the purported mistake. Things were as he predicted and  Wootters, Zurek and Dieks sent their refutations of the FLASH scheme to Nature and Physics Letters in August 1982. Peres concluded that Herbert's paper (plus his recommendation to publish it, I would add) was the spark needed to find the no-cloning theorem.

Peres wondered, in the quoted essay, given the importance and simplicity of the theorem, why it was not discovered fifty years earlier.  In the opinion of Scarani {\it et al.} ``there is no obvious answer to this question." \cite{scarani}  I cannot but agree with such a view but it is ironic that the first demonstration of the theorem \cite{park} had appeared 12 years earlier in the first issue of a new journal, Foundations of Physics, whose editors were Henry Margenau and Wolfgang Yourgrau. The  \textit{solo} paper entitled ``The concept of transition in quantum mechanics" was signed by James Park, a former student of  Margenau. Even more ironic is the fact that Foundations of Physics happens to be the same journal where the erroneous FLASH paper was published 12 years later.

It is believed that somehow everybody secretly knew that quantum information cannot be copied. An example of this belief is Peres' dictum ``these things were known to those that know things well" \cite{Ppaper} or the recent, partly humoristic, assertion by computer scientist Scott Aaronson in his blog \cite{Ablog} ``if you want to become immortal, just find some fact that everyone already knows and give it a name!" Aaronson refers to a paper by Stephen Wiesner \cite{Wpaper} that was published in 1983 although apparently a draft was written around 1970. Wiesner presented some quantum cryptographic schemes in which the impossibility of quantum copying is implicit. Recently, in the same spirit, Zeilinger \cite{Zpaper} mentions that ``... it is well known in the community that Wiesner had these ideas already in the 1970s, communicating them verbally at least to Charles Benett, but being unable to get a paper published." Scarani {\it et al.} mention in their authoritative review on quantum cloning \cite{scarani} some missed oportunities to prove the theorem. They specifically mention Townes, in relation to phenomenological equations describing the maser and Wigner, in relation to biological cloning. 

The  Letter to Nature written by Wootters and Zurek has been cited to date \cite{wok} over 2000 times, Dieks' paper over 650, while Park's article has received only 11 citations, \cite{fopw} none of which refers to his proof of the no-cloning theorem. Thus, my first goal here is contributing to give Park the credit he deserves. Also, I will elaborate on the reasons why Park's paper was virtually overlooked and why today it remains largely unknown. 

In Sect. II a brief summary is given of Wiesner's quantum cryptography paper and Herbert's FLASH paper. The well-known versions of the theorem are revised in Sect. III. A discussion of Park's paper along with a few remarks summarizing his particular views on quantum mechanics, are given in Sections IV and V. In Sect. VI, I present my conclusions.

\section{Previous ideas related to no-cloning}

\subsection{Wiesner's quantum money paper}
Wiesner \cite{Wpaper} proposed a method to create money that cannot be counterfeit. \cite{aaronson0} Weisner's idea was to include in each bill a bunch of qubits, ``isolated two-state physical systems such as, for example, isolated nuclei of spin 1/2." \cite{Wpaper} Then, ``The person holding the bill would not know the spin axis of each of the qubits. The quantum no-cloning theorem guarantees that if the bill holder tries to copy it, he damages it and cannot end up with two good bills." \cite{farhi} Actually Wiesner related his scheme to ``restrictions on measurement related to the uncertainty principal [sic]." Wiesner argues that there is no way of duplicating the money because ``if one copy can be made then many copies can be made by making copies of copies." This procedure would lead to an unlimited supply of systems in the same state. Wiesner continues ``the state could be determined and the sequence recovered. But this is impossible."

Wiesner's paper is certainly important and, according to many, it laid the foundation of quantum cryptography, but an explicit proof of the impossibility for copying unknown quantum information is not provided, although it is implicit in his schemes. The fact of that impossibility is implicit in Wiesner's schemes, but according to Peres' account \cite{Ppaper} these ideas did not influence the path to finding the no-cloning theorem. In fact, Wiesner \cite{Wpaper} managed to get his paper published only in 1983 after Wootters and Zurek \cite{WZpaper} and Dieks \cite{Dpaper} published their papers.

\subsection{Herbert's {\rm FLASH} paper}
This paper \cite{Fpaper} is unusual in many respects: i) The author was not afiliated with any academic institution (his address was given as Box 261, Boulder Creek, California); ii) Esalen Institute \cite{esalen}  is acknowledged. Esalen became the center of the New Age movement, among other counterculture alternative trends; iii) the title contains the acronym FLASH, which stands for the strange ``First Laser-Amplified Superluminal Hookup"; iv) the author recognizes that his purpose is designing a system that permits faster-than-light signaling; and finally v) a new kind of measurement is introduced, the Third Kind, named after Pauli's First and Second Kind measurements. A paper containing such surprising claims belongs to the class that would be rejected right away by most editors. Some courage was probably needed to send the paper to reputable referees like Peres and Ghirardi who quicky realized that the paper was erroneous, but Peres was even more courageous to accept the paper, seeing its potential influence for further research.

Essentially, the question that Herbert wonders about is: can ``quantum connectedness act as a medium for superluminal communication?" To answer the question he proposes an experiment inspired in Einstein-Podolsky-Rosen (EPR), \cite{epr} but knowing that the statistics of the measurements at one end of an EPR setup are independent of the measurements made at the other end, he proposes to make many copies of the photon at one end after a given measurement of the entangled photon at the other end has been made. With this so-called ``measurement of the Third Kind" an experimenter could distinguish instantly what single measurement has been made at a location separated from him by a space-like interval. 

On the other hand, the FLASH paper is one of the reasons for the title of the book ``How the hippies saved physics" \cite{davidkaser} written by the historian of science David Kaiser. This book explains the creation in 1975 of the informal ``Fundamental Fysiks Group" by Elizabeth Rauscher and George Weismann, graduate students at Berkeley at the time. Kaiser argues that ``The hippie physicists' concerted push on Bell's theorem and quantum entanglement  instigated major breakthroughs... The most important became known as the `no-cloning theorem'." Herbert was one of these hippies associated to the ``Fundamental Fysiks Group." Kaiser goes farther to assert: ``The no-cloning theorem emerged directly from the fundamental Fysiks Group's tireless efforts ... to explore whether Bell's theorem and quantum entanglement might unlock the secrets of mental telepathy and extrasensory perception...". Kaiser aligned with Peres in giving Herbert a fundamental role in the genesis of the no-cloning theorem. 

\section{The 1982 discoveries of the theorem}

\subsection{Wootters and Zurek: A single quantum cannot be cloned}
The first sentence of the Letter \cite{WZpaper} is: ``Note that if photons could be cloned, a plausible argument could be made for the possibility of faster-than-light communication [2]," where [2] refers to Herbert's paper. And later ``The actual impossibility of cloning photons, shown below, thus prohibits superluminar communication by this scheme." The proof shows that if a device is able to copy an incoming photon with vertical or horizontal polarization, it will be unable to copy a photon described by a pure state given by any arbitrary linear combination of the two polarization states. A cloning machine would have the following effect on a photon with polarization state $|s\rangle$:

\begin{equation}
|A_{0}\rangle |s\rangle \rightarrow |A_{s}\rangle |ss\rangle \;,
\label{eqz1}
\end{equation}

\noindent
where $|A_{0}\rangle$ and $|A_{s}\rangle$ are the initial and final states of the machine. The symbol $|ss\rangle$ refers to the state of two photons with polarization $|s\rangle$. If a transformation like the one represented in Eq. (\ref{eqz1}) can be accomplished for a photon with vertical polarization $|\updownarrow\rangle$ and for a photon with horizontal polarization 
$|\leftrightarrow\rangle$, its effect on a state given by an arbitrary linear combination of the two polarizations states will be

\begin{equation}
|A_{0}\rangle( \alpha|\updownarrow \rangle+\beta |\leftrightarrow \rangle) \rightarrow \alpha|A_{\rm vert}\rangle|\updownarrow\rangle+\beta|A_{\rm hor}\rangle |\leftrightarrow\rangle \;.
\label{eqz2}
\end{equation}

\noindent
If the apparatus states $|A_{\rm vert}\rangle$ and $|A_{\rm hor}\rangle$ are not identical, then the two photons emerging from the apparatus are in a mixed state of polarization. If these apparatus states are identical, then the two photons are in the pure state 

\begin{equation}
|\phi\rangle=\alpha|\updownarrow \updownarrow \rangle+\beta |\leftrightarrow \leftrightarrow \rangle \;.
\label{eqwz1}
\end{equation}

\noindent
This state is not the same as the pure state which would be obtained by copying the original photon:

\begin{equation}
|\phi'\rangle=\alpha^{2}|\updownarrow \updownarrow \rangle+\beta^{2} |\leftrightarrow \leftrightarrow \rangle+2^{1/2} \alpha \beta |\updownarrow \leftrightarrow \rangle;.
\label{eqwz2}
\end{equation}

Thus, a cloning machine cannot exist for arbitrary states, or equivalently it cannot clone unknown states.

\subsection{Dieks: Communication by EPR devices}
The received date of Dieks' paper in Physics Letters \cite{Dpaper} was only six days later than the received date for the Letter to Nature of Wootters and Zurek. Dieks analyzes Herbert's FLASH proposal  in connection to Bohm's version of EPR-type experiments. The FLASH idea requires the existence of a multiplying device able to generate many copies of a state. Dieks showed that such a device does not comply with the laws of quantum mechanics.

While Wootters and Zurek used photons for their proof, Dieks' proof deals with electons. Dieks analyzes an EPR-like experiment in which a state with spin zero decays into two spin 1/2 electrons. An experimenter {\bf A} can choose to measure the $x$-component or the $z$-component of the spin of electron I. Electron II enters a multypling device after {\bf A} has performed her measurement upon I. The device produces a burst of electrons in the same spin state as electron II. The large number $N$ of electrons coming from the multiplier are examined by an observer {\bf B} by means of a Stern-Gerlach apparatus adjusted to measure $s_x$. There are two possibilities: (i) Observer {\bf A} measures $s_x$, then electron II is in an eigenstate of $s_x$. Subsequent measurements made by observer {\bf B} will find that {\it all} $N$ electrons gives $s_{x}=1/2$ or $s_{x}=-1/2$, (ii) Observer {\bf A} measures $s_z$. The electrons emerging from the multiplier will be in an eigenstate of $s_z$. Measurements made by observer {\bf B} will find that, for $N/2$ electrons,  $s_x=1/2$, and for the other $N/2$ electrons,  $s_x=-1/2$. Therefore, observer {\bf B} would know which measurement observer {\bf A} performed.

This process can be represented simbolically as follows:

\begin{equation}
|x_{\pm} \rangle |M_{0} \rangle \rightarrow |M_{\pm}\rangle |x_{\pm};N\rangle \;,
\label{eq31}
\end{equation}

\noindent
where ``$|M_0\rangle$ is the `neutral' state of the multiplier before the electron enters; $|x_{\pm};N\rangle$ represents the $N$-particle state of $N$ electrons all in the same spin eigenstate 
$|x_{\pm}\rangle$; $|M_{\pm}\rangle$ is the state in which the multiplier is left." If the $N$ cloned electrons interact with a Stern-Gerlach apparatus all of them will arrive in either the $s_{x}=+1/2$ or $s_{x}=-1/2$ channel, depending on whether the incoming electron was in spin state $|x_{+}\rangle$ or $|x_{-}\rangle$. 
But, if the incoming electron is in one eigenstate of the operator representing the spin projection along the $z$ axis, $s_z$, $|z_{\pm}\rangle=\frac{\sqrt{2}}{2}(|x_{+}\rangle \pm |x_{-}\rangle)$, its effect on the multiplying device is completely determined, due to the linearity of the quantum mechanical evolution operator, and can be written simbolically as:

\begin{equation}
|z_{\pm} \rangle |M_{0}\rangle \rightarrow  \frac{\sqrt{2}}{2} (|M_{+}\rangle |x_{+};N\rangle \pm |M_{-}\rangle |x_{-};N\rangle ) \;.
\label{eq32}
\end{equation}

\noindent
However, Dieks continues, the definition of the multiplier,  as employed in the FLASH scheme, would require that the cloned electrons were in the state $z_{\pm}$:

\begin{equation}
|z_{\pm}\rangle |M_{0}\rangle \rightarrow |M_{\pm}\rangle |z_{\pm};N\rangle \;,
\label{eq33}
\end{equation}

\noindent
where the right-hand member represents a burst of electrons all in the $|z_{\pm}\rangle$ state. This final state is not identical to the state predicted by quantum mechanics, given in the right-hand side of Eq. (\ref{eq32}). This is an entangled state of the $N$-electron-apparatus system, while the state in Eq. (\ref{eq33}) is a product of an apparatus state and an $N$-electron state. Therefore, the FLASH proposal for superluminal communication is not consistent with quantum mechanics.

\section{Park's proof of the no-cloning theorem}

Park presents in Section 3 of \cite{park} a measurement scheme which 
contains a demonstration of the no-cloning theorem.  His model analyzes the interaction between two spins {\bf S} and {\bf M}. Each spin is a two-state system, so the combined {\bf S}+{\bf M} system has a four-dimensional tensor product space. Park studies whether a measurement scheme is possible based on the existence of a unitary evolution operator $T$ such that it effects the following state evolution for {\bf S}+{\bf M}

\begin{equation}
T |\phi\rangle |\alpha\rangle=|\phi\rangle  |\phi\rangle \;,
\label{eq41}
\end{equation}

\noindent
where $|\phi\rangle=a|\alpha\rangle+b|\beta\rangle$, with $|\alpha\rangle$, $|\beta\rangle$  eigenvectors of the operator $s_z$ defined in Sect. III.B.  
If this interaction exists, it ``transfers the state specification of {\bf S} to {\bf M}, yet {\bf S} emerges in the same state...Hence, measurements upon {\bf M} yield measurement results for {\bf S} without changing the state of {\bf S}." In other words, if the system in the unknown state could be cloned, measurements made upon these clones would reveal the premeasurement state without disturbing the original system. Thus, a nondisturbing $T$ must satisfy for all $a$, $b$ values

\begin{equation}
T (a |\alpha\rangle+b |\beta\rangle)  |\alpha\rangle=(a|\alpha\rangle+b|\beta\rangle)  (a|\alpha\rangle+b|\beta\rangle) \;.
\label{eq42}
\end{equation}

\noindent
Invoking the linearity of $T$ and expanding, the previous expression becomes

\begin{equation}
a T |\alpha\rangle |\alpha\rangle+b T |\beta\rangle |\alpha\rangle=a^{2}|\alpha\rangle |\alpha\rangle +ba|\beta\rangle |\alpha\rangle+ab|\alpha\rangle|\beta\rangle+b^{2}|\beta\rangle
|\beta\rangle \;.
\label{eq43}
\end{equation}

\noindent
Next, Park proves that if such a $T$ exists it must depend on $a$ and $b$, and therefore on $|\phi\rangle$. For that,  he shows that if $T$ does not depend on $|\phi\rangle$ an absurd result is obtained: Considering the scalar product of Eq. (\ref{eq43}) with $\langle \alpha| \langle \alpha|$, we get

\begin{equation}
a\langle \alpha \alpha|T|\alpha\alpha\rangle+b\langle \alpha \alpha|T|\beta\alpha\rangle=a^{2}\;,
\label{eq44}
\end{equation}

\noindent
which must hold for all $a$ and $b$ values and $|a|^{2}+|b|^{2}=1$. Due to this arbitrariness, some matrix elements can be readily found by giving specific values to $a$ and $b$ in
Eq. (\ref{eq44}):

\begin{equation}
\langle \alpha \alpha|T|\alpha\alpha\rangle=1, \; {\rm from} \;\;  a=1, b=0 \;,
\label{eqad1}
\end{equation}

\begin{equation}
\langle \alpha \alpha|T|\beta\alpha\rangle=0, \; {\rm from} \;\; a=0, b=1 \;,
\label{eqad2}
\end{equation}

\begin{equation}
\frac{1}{\sqrt{2}} \langle \alpha \alpha|T|\alpha\alpha\rangle+ \frac{1}{\sqrt{2}} \langle \alpha \alpha|T|\beta\alpha\rangle)=\frac{1}{2}, \;{\rm from} \;\; a=b=\frac{1}{\sqrt{2}} \;. 
\label{eqad3}
\end{equation}

\noindent
Combining Eqs. (\ref{eqad1}-\ref{eqad3}),
the absurdity $1/\sqrt{2}=1/2$ is obtained.
Therefore, Park concludes that ``there exists no {\it simple} nondisturbing measurement between two spins." This is due to the fact that a generic unknown state cannot be copied, which is the no-cloning theorem!. 

\section{Some controversial views in Park's paper}

Park does not seem to have viewed the no-cloning result as a major aspect of his paper and does not refer to it in the abstract or conclusions. Essentially, Park's proof of the impossibility of cloning is a byproduct of his quest to prove that nondisturbing quantum measurements are possible. If cloning were allowed {\it simple} nondisturbing measurements could be performed. Note that Herbert's {\it impossible} ``Third Kind measurements" \cite{Fpaper} are nothing but nondisturbing measurements, in the sense that an unknown quantum state could be determined, without altering it, by making measurements upon clones.

From an orthodox point of view,  unavoidable perturbations in any quantum measurement project the state, $|\phi\rangle$, of the system being measured into the eigenstate 
$|\alpha_k\rangle$ corresponding to the measurement result $a_k$. 
This implies that  $|\langle \alpha_k, \phi\rangle|^2$ gives the probability for finding the system in state $|\alpha_k\rangle$. Park does not endorse this idea, as the following sentence shows: ``while it is factually correct that measurement operations upon microphysical systems tend to have catastrophic effects upon their states, the notion of uncontrollable disturbance of a state by a measurement act, ...should not be regarded as a {\it universal} trait of the measurement act." \cite{park}

In spite of having proved that cloning {\it unknown} states is not possible, Park perserverates in his search, asking whether a {\it specific} quantum state can be copied by an interaction especially designed for that state. He answers affirmatively and name such interactions ``nondisturbing measurement procedures of the historical type." Wootters and Zurek \cite{ WZpaper} explicitly commented on this issue: ``linearity does not forbid the amplification of any given state by a device designed especially for that state, but it does rule out the existence of a device capable of amplifying an arbitrary state."

In Ref. \cite{park} a very contrived example is given of a {\it historical} nondisturbing measurement. A specific spin state, $|\phi\rangle$, of a system {\bf S} is copied, by means of an unitary transformation, onto a system {\bf M}. Then, the claim is made that further measurements on {\bf M}, of a given observable, qualify as measurements of the same observable for {\bf S}. This idea is at odds with the orthodox interpretation of quantum mechanics because the particular outcome that takes place in a given quantum measurement is not decided before the measurement is made, unless $|\phi\rangle$ is an eigenstate of the observable being measured. Consequently, the value measured on {\bf M} will be different, in general, to the value measured on {\bf S}. On the other hand, the property measured on {\bf M} could correspond to an observable represented by an operator that does not commute with the operator for which $|\phi\rangle$ is an eigenstate.  If Park's claim was correct, one could ascertain the value of two incompatible observables, which we know to be against the conceptual core of orthodox quantum theory. 

The controversial meaning given to nondisturbing measurements corresponds to an unorthodox view on quantum mechanics, outside the mainstream, advocated by Park and his thesis advisor, Margenau. The title of Ref. \cite{new}  ``Simultaneous measurability in quantum theory," which contains part of Park's doctoral dissertation, is a explicit statement of intent. In the abstract of the paper it is stated that ``the much quoted `principle' of incompatibility of noncommuting observables is false." The same idea appears in Section III of Ref. \cite{park} where it is mentioned that the {\it historical} measurement procedure ``demolishes another quantum myth, namely, that if the value of one member of a noncommuting pair of observables is known, the other cannot be measured without destroying the certain value of the first."

It should be mentioned though that, for Park, the state vector refers irreducibly to an ensemble, ``...the quantum postulates seem to correlate the state concept to a system only through an intervening ensemble of such systems identically prepared." \cite{parkmeasure} Note that an ensemble may be an aggregation of elements all present at once, but, equally, it might refer to just one system prepared and sequantially reprepared.  Measurements of a certain observable upon enough members of an ensemble give a probability distribution.  In the ``historical nondisturbing procedure" analyzed above, measurement results of {\bf M}-observables yield {\it the same probability distributions} as measurement results of {\bf S}-observables would have given. Only in this sense both measurements are equivalent.

\section{Conclusions}

According to Kaiser \cite{davidkaser} the no-cloning theorem, so important in the development of quantum information theory, \cite{qin} was discovered thanks to the concerted efforts of some unusual physicists that decided to create a study group in 1975 to find esoteric applications of quantum mechanics. As a result of the ideas generated in the group a paper came out that was so wrong that a respected referee considered it worth publishing. This happened in 1982 and the journal was Foundations of Physics.  Two independent demonstrations of the no-cloning theorem were published as a reaction to the wrong paper.  

The irony of the previous story, that seems to be scrupulously accurate, is that a proof of the impossibility of copying unknown quantum states had been already published in the same journal, Foundations of Physics, 12 years earlier but nobody took notice of it. The fact that the emphasis of the paper is put on other aspects of quantum mechanics overshadows the mathematical proof of the theorem and probably is one of the reasons why the paper is not well known. Thus, I concur with Peres and Kaiser that it seems quite plausible that the ``spark" of a flagrantly wrong proposal (the FLASH paper) was needed to get the community to pay attention to the issue of copying quantum information.

Park sent his paper to Foundations of Physics in June 1969 and it appeared published in the first issue of the journal. The paper is not indexed in the Web of Knowledge \cite{wok} because only papers of the journal Foundations of Physics published after 1973 are indexed in that database. Nonetheless, 11 citations to Park's paper are registered. \cite{fopw}
Obviously none of these citations mentions the demonstration of the impossibility to copy quantum information. It is striking that one of the 11 citing papers is authored by Wootters and Peres among others, \cite{woot} and another one by Peres!. \cite{peres} These references to Park's work are given in very neutral sentences. For example: \cite{woot} ``...This spin-exchange measurement [4] illustrates an essential feature of quantum information: it can be swapped from a system to another, but it cannot be duplicated or cloned, [5]" where [4] is Park's article \cite{park} and [5] is the paper by Wootters and Zurek. \cite{Wpaper}  Interestingly, both articles are cited in the very same sentence; but the merit for the no-cloning theorem goes to Wootters and Zurek, while Park's paper is cited for containing an example of a spin-exchange measurement. On the other hand, it is remarkable that Park has not claimed a role in the discovery of the no-cloning theorem, or at least I have not been able to find any trace of complaint.

The editorial preface, written by Margenau and Yourgrau, appearing in the first issue of Foundations of Physics, states ``One wonders whether brilliant ideas are not lost by this restrictive attitude," referring to the fact that speculative research was not encouraged by the journals existing at that time. The case of the no-cloning theorem shows that brilliant ideas can be lost due to many kinds of restrictive attitudes, like the discounting of authors or works from outside the mainstream.  If Bohr and Von Neumann represent the quantum ortodoxy, Park should be labeled as a heretic. He, as Margenau, strongly opposed the projection postulate of Von Neumann and the Bohrian idea that the act of measurement disturbs the measured system. The fact that Park's motivation was the exploration of anti-Bohrian ideas may have contributed to the cold reception of his paper. 

Also, when Park wrote the article that contains the no-cloning theorem, research in the foundations of quantum mechanics was discouraged. The prevalent attitude at the time is well summarized by Mermin's slogan ``shut up and calculate." \cite{mermin}
But things were not very different when Wootters and Zurek or Dieks published their papers. Ten years after publication, the Letter to Nature only had been cited 29 times, and Dieks' paper had received 11 citations. It was only in 1998, 15 years after publication of those versions of the theorem and 28 years after Park's original proof, that the theorem started to be profusely cited.  By 1998 other fundamental papers had been published contributing to the new field of quantum information. \cite{woot, bennett, zeilinger2} But according to Scarani {\it et al.}, \cite{scarani} the real trigger of the explosion in reseach related to quantum cloning was the publication of a paper by Buzek and Hillery \cite{buzek} suggesting the possibility of imperfect cloning. This may show that sometimes brilliant ideas can be lost or forgotten because they are ahead of their time.

\end{document}